\documentclass[aps,twocolumn,preprintnumbers,showpacs,showkeys,nofootinbib%
]{revtex4}
\usepackage{epsfig}
\usepackage{amssymb,amsmath,amsfonts,amsthm,graphicx,psfrag}
\graphicspath{{./Figures/}}                 
\usepackage{array}
\usepackage{picture}
\usepackage{subfig}
\usepackage{graphicx}
\usepackage{textpos}
\usepackage{color}

\newcommand{\beq}{\begin{equation}}
\newcommand{\eeq}{\end{equation}}
\newcommand{\bea}{\begin{eqnarray}}
\newcommand{\eea}{\end{eqnarray}}
\newcommand{\aabf}{{\mathbf{A}}}

\newcommand{\re}{\operatorname{\mathfrak{Re}}}
\newcommand{\tr}{\operatorname{T\!r}}

\newcommand{\ds}{\displaystyle}

\newcommand{\aasymt}{{\cal A}}

\begin{document}

\title{Studies of Correlations in the Critical Domain of the $SU(2)$ Gluodynamics.}

\author{R.~N.~Rogalyov}
\affiliation{Institute for High Energy Physics NRC ``Kurchatov Institute'', 142281 Protvino, Russia }
\author{V.~G.~Bornyakov}
\affiliation{Institute for High Energy Physics NRC ``Kurchatov Institute'', 142281 Protvino, Russia}

\author{N.~V.~Gerasimeniuk}
\affiliation{Pacific Quantum Center, Far Eastern Federal University, 690922 Vladivostok, Russia}

\author{E.~A.~Kozlovsky}
\affiliation{Institute for High Energy Physics NRC ``Kurchatov Institute'', 142281 Protvino, Russia}

\author{V.~K.~Mitjushkin}
\affiliation{Joint Institute for Nuclear Research, 141980 Dubna, Russia}


\begin{abstract}
By considering the example of $SU(2)$ gluodynamics,
we check numerically the idea that 
the strong correlation of the Polyakov loop 
with the longitudinal gluon propagator 
and related quantities
can be used to substantially reduce the finite-volume effects as well as for 
extrapolation in temperature in the critical domain.
\end{abstract}

\keywords{Lattice gauge theory, gluon propagator, }

\pacs{11.15.Ha, 12.38.Gc, 12.38.Aw}

\maketitle

\section{Introduction}
\label{sec:introduction}

Recently it was found 
\cite{Bornyakov:2018mmf,Bornyakov:2021pls}
that a significant correlation
of the Polyakov loop with the low-momentum
gluon propagator as well as with the asymmetry 
between the dimension-2 chrmoelectric and chromomagnetic condensates makes it possible to 
adequately describe the critical behavior of these 
quantities and find the critical exponents on the basis of the universality hypothesis and some plausible assumptions.

Here we study some consequences of the observed correlations in more detail.

We check numerically the idea outlined in Section~\ref{sec:data_transform}
how to use the correlation between the Polyakov loop and 
the above-mentioned quantities in order to reduce the finite volume effects. Roughly speaking, we use temperature fluctuations, which are inevitable in a finite-volume system, in order to "scatter" the data 
for the quantities under study over a finite temperature interval. Considering that the quantities 
under study depend strongly on the Polyakov loop
and only weakly on the temperature itself,
temperature fluctuations has only a little (negligible) effect on these quantities provided that the Polyakov loop is fixed. Thus we choose the fluctuation 
such that each particular configuration
gets the temperature connected with the Polyakov loop
for this configuration by formula (\ref{eq:crit_index_beta}), see 
 Section~\ref{sec:data_transform}.
We argue that this procedure makes it possible 
to approach the infinite-volume limit
for the quantities correlated with the Polyakov loop
even though the initial data set is obtained on a small  lattice.

A byproduct of the proposed improvement  
of the data set is the possibility to extract information 
about the quantities under study at a given temperature 
from the data obtained at another (albeit close) 
temperature. An example of such extrapolation through 
the critical temperature is considered.

\section{Definitions and simulation details}

We study $SU(2)$  lattice gauge theory with 
the standard Wilson action in the Landau gauge. 

The chromo-electric-magnetic 
asymmetry introduced in \cite{Chernodub:2008kf} and studied in
\cite{Bornyakov:2016geh} is defined as
\beq
 {\cal A} \equiv (\big\langle A_E^2 \big\rangle - \frac{1}{3} \big\langle A_M^2 \big\rangle)/T^2 \; ,
\label{asymmetry}
\eeq
where
\bea
\big\langle A_E^2 \big\rangle &=& g^2 \big\langle  A^a_4 (x)  A^a_4 (x) \big\rangle, \\ \nonumber
\big\langle A_M^2 \big\rangle &=& g^2 \big\langle  A^a_i (x)  A^a_i (x) \big\rangle. \nonumber
\eea

The lattice link variable $U_{x\mu} \in SU(2) $ is related to the Yang-Mills 
vector potential $A_\mu^b(\vec x, x_4)$ as follows.
One determines a Hermitian traceless matrix 
\beq
z={1\over 2\imath}\left( U_{x\mu}-U_{x\mu}^\dagger  \right)
\eeq
which is connected with the dimensionless vector potentials ($a$ is the lattice spacing)
\beq
u^b_\mu(x) = \;-\; {ga\over 2} A_\mu^b(x)\;  ,
\eeq
by the formulas
\beq
z_{ij}= u^b_\mu(x) \Gamma^b_{ij}, \qquad u^b_\mu(x) = 2\,\mathrm{T\!r} \Big(\Gamma^b z\Big) = 2 \Gamma^b_{ij} z_{ji}\,,
\eeq 
where $\Gamma^a$ are Hermitian generators of $SU(2)$ 
normalized so that
\beq
\langle \Gamma^a \Gamma^b \rangle \equiv \mathrm{T\!r} (\Gamma^a \Gamma^b) = \Gamma^a_{ij} \Gamma^b_{ji} = {1\over 2} \delta^{ab}
\eeq

In the fundamental representation,
\begin{displaymath}
\Gamma^a = \begin{array}{lcl}
                     \ds {\sigma^a \over 2}    & \mbox{~~for} & SU(2) 
                    \end{array}.
\end{displaymath}

Transformation of the link variables $U_{x\mu}$ 
under gauge transformations $g_x$ has the form 
$$ U_{x\mu}
\stackrel{g}{\mapsto} U_{x\mu}^{g} = g_x^{\dagger} U_{x\mu} g_{x+\mu}\;.
$$

The lattice Landau gauge condition is given by
\beq
(\partial \aabf)_{x} = \sum_{\mu=1}^4 \left( \aabf_{x\mu}
- \aabf_{x-\hat{\mu};\mu} \right)  = 0 \,.
\eeq
It represents a stationarity condition for the gauge-fixing functional
\beq\label{eq:gaugefunctional}
F_U(g) = \frac{1}{4V}\sum_{x,\mu}~\frac{1}{3}~\re\tr~U^{g}_{x\mu} \;,
\eeq
with respect to gauge transformations $g_x~$.

\vspace*{2mm}
Another quantity of interest studied here is the longitudinal gluon 
propagator at zero momentum $D_L(0)$. Its definition can be found, e.g. in
\cite{Bornyakov:2016geh}.

Our calculations are performed on asymmetric lattices $N_t\times N_s^3$,
where $N_t$ is the number of sites in the temporal direction. In our study, 
$N_t=8$ and $N_s$ varies so that $L=N_sa\approx 2.6$~fm or 6.0~fm.
The physical momenta $p$ are given by $\hat p_i=\big(2/a\big) \sin{(\pi
k_i/N_s)}, ~~\hat p_{4}=(2/a) \sin{(\pi k_4/N_t)}, ~~k_i \in (-N_s/2,N_s/2], k_4 \in
(-N_t/2,N_t/2]$. We consider only soft modes $p_4=0$. 

The temperature $T$ is given by $~T=1/aN_t~$ where $a$
is the lattice spacing determined by the coupling constant. We use the parameter
\beq
\tau = {T-T_c \over T_c}
\eeq 
at temperatures close to $T_c$ and call it "temperature" where it is possible. We also use the letter $t$ for the designation of this parameter
when different values of this parameter are involved in our speculations.

We provide information on lattice spacings, temperatures, and other parameters used in this work
in Table~\ref{tab:statistics_SU2}.

\begin{table}[tbh]
\begin{center}
\vspace*{0.2cm}
\begin{tabular}{|c|c|c|c|c|c|} \hline
           &          &               &                &     &      \\[-2mm]
 ~~~~$4/g^2$~~~~ & ~~$a$,~fm~ & $a^{-1}$,~GeV &  ~~~~$\tau$~~~~& $L$, fm &  $L$, fm\\
           &          &               &                &  ($N_s=32)$      &   ($N_s=72$) \\
   \hline\hline
2.510 & 0.0831 & 2.374 & -0.0013 & 2.66 & 5.98   \\
2.511 & 0.0826 & 2.389 & 0.0019 & 2.65 & 5.96 \\
2.512 & 0.0826 & 2.389 & 0.0051 & 2.64 & 5.95 \\
2.513 & 0.0823 & 2.397 & 0.0083 & 2.64 & 5.93 \\
2.515 & 0.0818 & 2.412 & 0.0148 &  2.62  & 5.89 \\
2.518 & 0.0810 & 2.436 & 0.0246 & 2.60 & 5.83 \\
2.521 & 0.0802 & 2.459 & 0.0345 & 2.57 & 5.77 \\
2.527 & 0.0787 & 2.507 & 0.0545 & 2.52 & 5.67 \\
\hline
\end{tabular}
\end{center}
\caption{Parameters associated with lattices under study for $SU(2)$ gluodynamics. 
}
\label{tab:statistics_SU2}
\end{table}

\begin{table}[tbh]
\begin{center}
\vspace*{0.2cm}
\begin{tabular}{|c|c|c|c|} \hline
     &      &               &             \\
~~~~$\tau$~~~~& ~~~~$N_s=32$~~~~  & ~~~~$N_s=72$~~~~ &  ~~~~$N_s=32$~~~~    \\
     &     &                &   improved          \\
   \hline\hline
 0.0019 & 2.314(29) & 2.947(68) &  3.270(11)   \\
 0.0051 & 2.303(30) & 2.776(24) &  2.8035(91) \\
 0.0083 & 2.194(29) & 2.629(26) &  2.577(10)  \\
 0.0148 & 2.445(31) & 2.326(25) &  2.216(10) \\
 0.0246 & 1.913(30) & 1.971(18) &  1.8801(87) \\
 0.0545 & 1.474(24) & 1.437(14) &  1.3043(69) \\
\hline\hline
\end{tabular}
\end{center}
\caption{Average values of the asymmetry. 
}
\label{tab:asymmetry_SU2}
\end{table}

\section{Description of Data Transformation}
\label{sec:data_transform}

First, we define the quantities under consideration.
Let $\mathbf{p}(\tau,V) $ be the function
describing the temperature and volume dependence 
of the average value of the Polyakov loop,
\beq
\langle {\cal P} \rangle =  \mathbf{p}(\tau,V) \equiv
\mathbf{p}_V(\tau) \;. 
\eeq
Asymptotic  behavior of this function 
in the infinite-volume limit has the form
\beq\label{eq:crit_index_beta}
\mathbf{p}_V(\tau) = C\tau^b + \underline{O}(\tau^{b+\omega\nu}), \quad \tau\to 0_+
\eeq
where $b\approx 0.326419$ \cite{Svetitsky:1982gs,Kos:2016ysd}, for more details see \cite{Engels:1998nv}. We use the letter $b$ instead of the conventional $\beta$ to avoid confusion with lattice parameter $\ds \beta={4\over g^2}$.

Now we define two functions. 
The former is the expectation value of the asymmetry,
\beq\label{eq:immed_aver}
f_0(t,V)=E_{t,V}(\aasymt)
\eeq 
where the subscript $(t,V)$ indicates that 
the expectation value is estimated
using the ensemble simulated on a lattice
characterized by reduced temperature $t$ and volume $V$.
The latter function is the conditional expectation value of the asymmetry at a given value of the Polyakov loop
\beq\label{eq:improved_asym}
f_{t,V}(\tau)=E_{t,V}\Big(\aasymt|{\cal P}=\mathbf{p}_V(\tau)\Big),
\eeq 
the Polyakov-loop value being associated with the temperature using the leading term on the r.h.s. of formula (\ref{eq:crit_index_beta}).

Correlation of the asymmetry with the Polyakov loop
and our knowledge of the temperature dependence of the Polyakov loop in the infinite-volume limit 
provide information that we use in the following improvement procedure.

We take the ensemble of configurations
simulated at temperature $t_0$ and volume $V$,
and find ${\cal P}$ and $\aasymt$ for each configuration. Then we assume that the configuration with the Polyakov loop value ${\cal P}$
should be associated with the temperature
\beq\label{eq:shifted_temperature}
\tau=\mathbf{p}_V^{-1}({\cal P}) \approx \left({{\cal P} \over C}\right)^{1/b}
\eeq 
rather than with $t_0$.
Thus we obtain the ensemble in which
temperature fluctuations are determined
by the Polyakov-loop fluctuations 
of the initial ensemble,
we call it the improved ensemble.
The scatter-plot in the ${\cal P}-{\cal A}$
plane obtained for the initial ensemble 
(see, e.g., \cite{Bornyakov:2018mmf,Bornyakov:2021pls})
is transformed into the scatter plot 
in the ${\cal A}-\tau$ plane.
In performing the above procedure, the points distributed in the 
${\cal P}-\tau$ plane get into the graph of the function
$\mathbf{p}_\infty(\tau)$.

\begin{figure}[htb]
\vspace*{-16mm}
\hspace*{-5mm}\includegraphics[width=9cm]{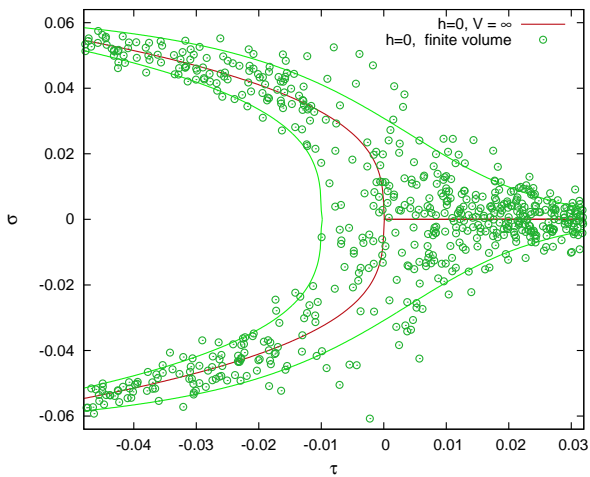}
\vspace*{-35mm}
\caption{Scatter plot in the temperature-magnetization plane.}
\label{fig:illustration}
\end{figure}

Similar averages can be introduced not only for ${\cal A}$ but also for the quantity ${\cal D} = ln\Big[ D_L(0)\sigma \Big]$, where 
$\sqrt{\sigma}=440$~MeV is the string tension (not to be confused with the magnetization in Fig.\ref{fig:illustration}).

For a clearer explanation of our main assumption,
let us consider a finite-volume ferromagnet 
as an example. One measures its magnetization $\sigma$
and temperature $\tau$ and marks each measurement by
the respective point in the $\tau-\sigma$ plane
(see green circles in Fig.\ref{fig:illustration}).
However, in the infinite-volume limit,
the distribution of such points shrinks 
to the red line corresponding to 
the critical behavior of the magnetization. 
Let us now consider some quantity $Q$ 
having a strong correlation with the magnetization
(so that its temperature dependence 
is determined mainly by the magnetization) 
and then  shift each
point in the $\tau-\sigma-Q$ 
space along the abscissa $(\tau)$ axis 
so that its projection on the $\tau-\sigma$ plane  
gets into the red line.
This results in only a little change of $Q$
because the magnetization remains unchanged.
Now one is tempted to consider that the shifted 
(improved) data points make it possible 
to estimate $Q(\tau)$ in the infinite-volume limit 
using the data obtained in a relatively small volume.
The additional information needed for such an estimate 
is provided by the knowledge of the infinite-volume
behavior of the magnetization and the correlation 
between $Q$ and $\sigma$.


Since ${\cal P}$ and $-{\cal P}$ give the same 
temperature, the positive Polyakov-loop sector 
and the negative Polyakov loop sector 
should be treated separately - the more so
the gluon propagators  behave differently 
in different center sectors \cite{Silva:2016onh}.

However, here we restrict our attention 
to the positive Polyakov-loop sector.


\section{Decrease of finite-volume effects}

\begin{figure*}[bht]
\vspace*{-16mm}
\hspace*{-5mm}\includegraphics[width=9cm]{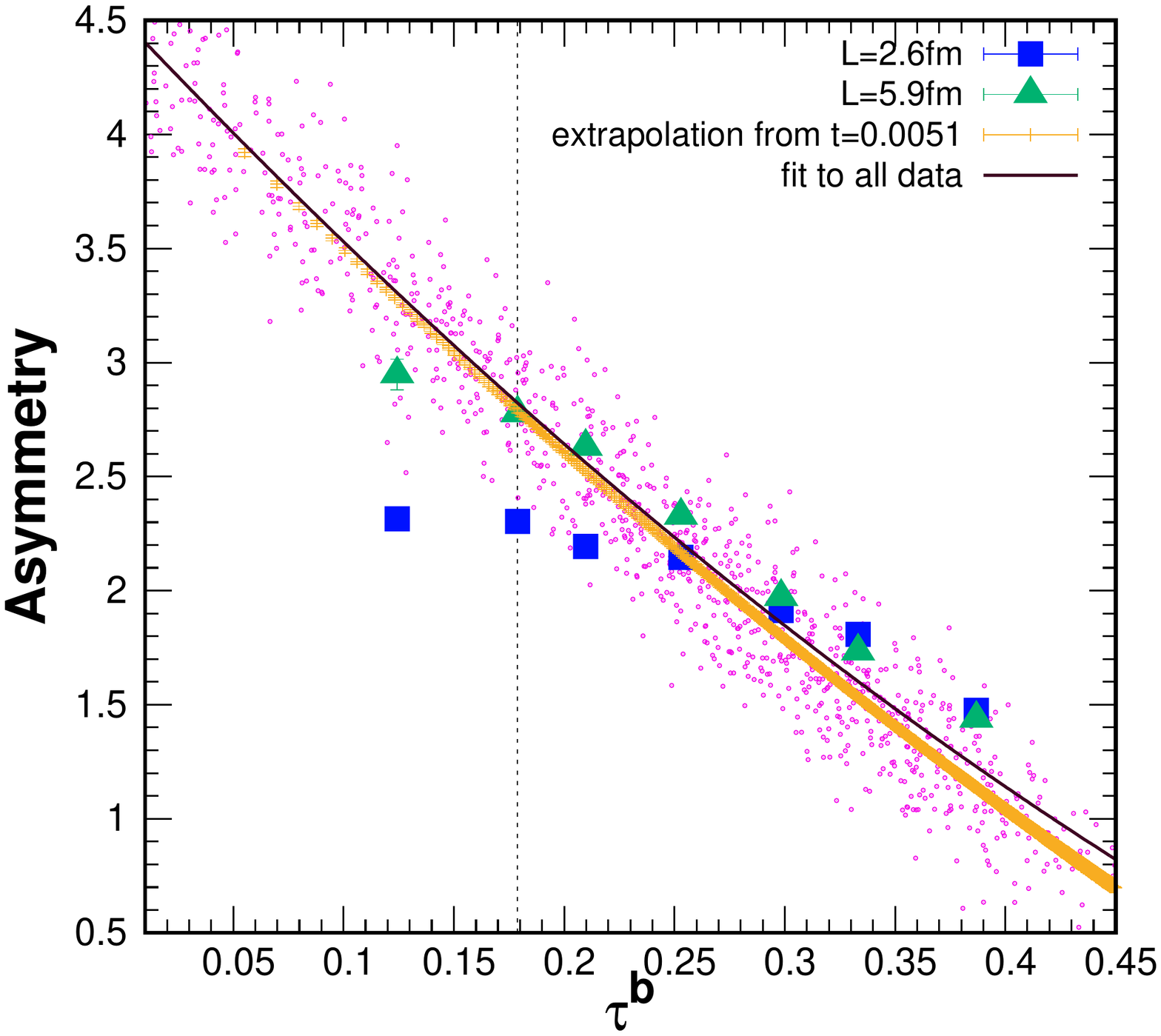}\hspace*{-8mm}
\includegraphics[width=9cm]{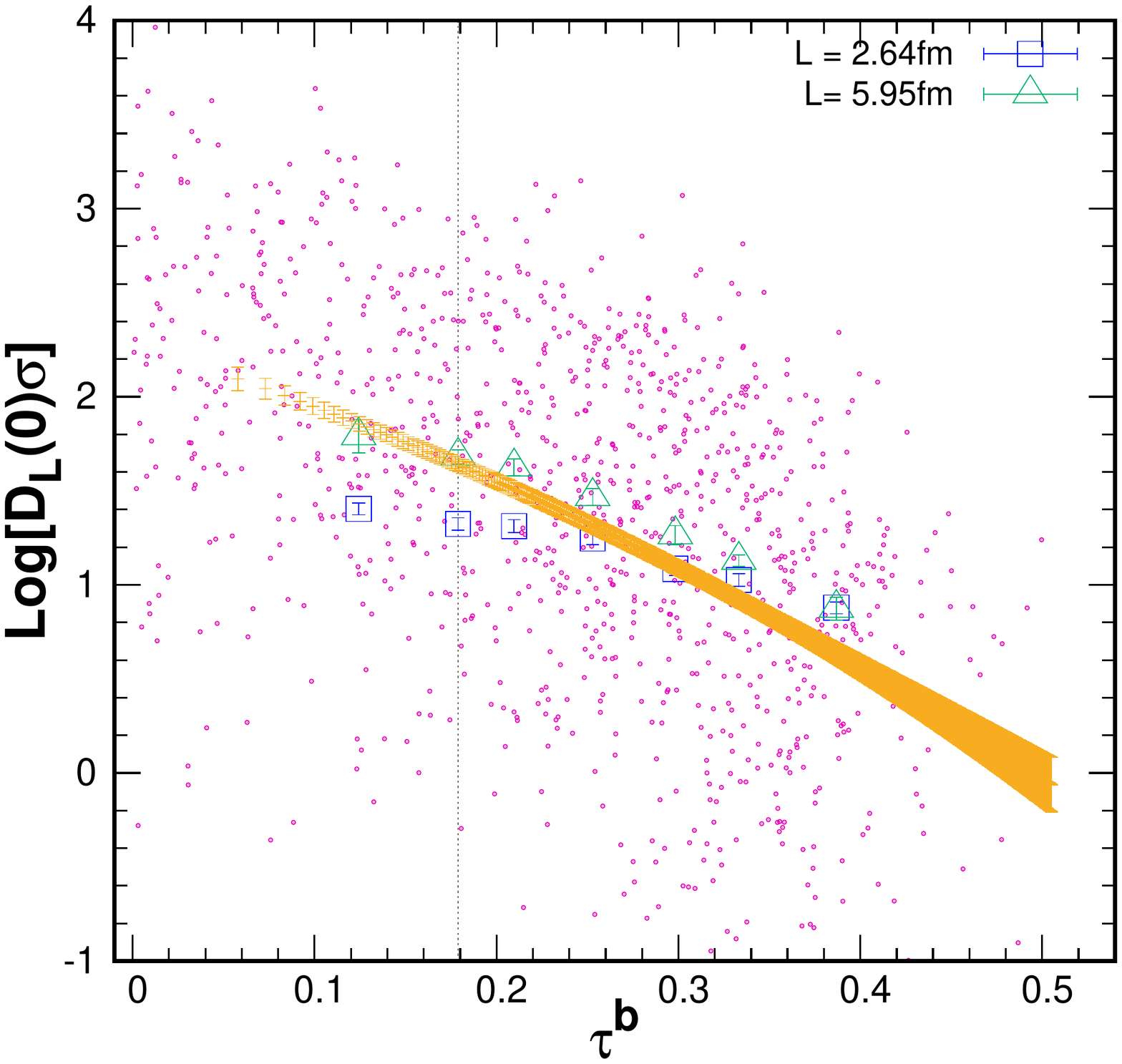}
\vspace*{-21mm}
\caption{The results for the improvement procedure
(yellow strips) are compared with the conventional averages obtained on the lattices of indicated sizes:
squares are the small-size lattices and triangles 
--- large-size lattices. Note that 
the extrapolation is performed 
using the data for smaller lattices. 
Vertical lines indicate the temperature from which 
the extrapolation was performed.}
\label{fig:FVE}
\end{figure*}

We rearrange the data obtained on 
a particular lattice (say, on that with 
$t=0.0051$ and $L=2.6fm$)
using our improvement procedure
in order to obtain the scatter plot 
in the asymmetry-temperature plane
shown on the left panel of Fig.\ref{fig:FVE}.
Then we perform regression analysis 
and employ the bootstrap technique
in order to find the dependence of the 
average value of the asymmetry 
as the function of the temperature.
The respective confidence corridor
is shown by yellow strip.

Thus we perform extrapolation from a particular value 
of the temperature used in simulations
to the range of temperature fluctuations
associated with the Polyakov-loop fluctuations
in the lattice volume under consideration.
It is seen in Fig.\ref{fig:FVE} that,
in the critical domain, the range of 
temperature fluctuations is rather wide.

It is clearly seen that, at sufficiently
small $\tau$ ($\tau\lesssim 0.02$), the improved data
and, therefore, the improved 
average of the asymmetry
is shifted to the values
associated with substantially greater volumes.
The significant deviation of the infinite-volume
values of the asymmetry
at $\tau\gtrsim 0.02$ can be explained by 
increasing role of the terms omitted in the approximate formula (\ref{eq:crit_index_beta}).

A similar procedure was performed with the data
for the zero-momentum longitudinal propagator.
We study the quantity ${\cal D}(\tau)= ln\Big[ D_L(0)\sigma \Big]$ because the scatter in the data for 
$D_L(0)$ itself is so wide that it hinders statistical analysis (the width of the distribution in $D_L(0)$ depends severely on the temperature).

The results are shown on the right panel of 
Fig.\ref{fig:FVE}, we see the regularity
similar to the case of the asymmetry.
However, the difference between the asymmetry and the 
propagator is that the fraction of variance unexplained
is much greater in the latter case.

In the case of asymmetry we also show the
regression curve obtained from the 
regression with the bootstrap procedure 
for the combined data set 
(the data simulated at different lattices are mixed together). Justification of this operation is discussed below.

\begin{figure}[htb]
\vspace*{-16mm}
\hspace*{-5mm}\includegraphics[width=9cm]{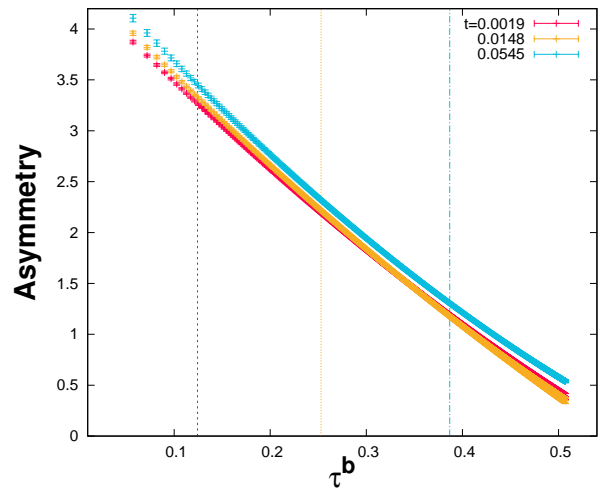}
\vspace*{-35mm}
\caption{Confidence corridors for the extrapolations
of the asymmetry from the data sets corresponding to the vertical dotted lines.}
\label{fig:corridors}
\end{figure}

The confidence corridors for $f_1(t;\tau,V)$ (see 
formula (\ref{eq:improved_asym}) for a definition) at 
$t=0.0019$, $t=0.0148$ and $t=0.0545$ and $V=(2.6$~fm$)^3$
are shown in Fig.\ref{fig:corridors}.
It is seen that the corridors 
for $t=0.0019$ and $t=0.0148$ are close to coincidence,
whereas the corridor obtained for the 
$t=0.0545$ data deviates from the others.
The confidence corridors for the other values of $t$
are not shown because it is difficult to present them 
in a common plot. However, our analysis reveals that
the confidence corridors of the asymmetry values 
are substantially overlap for $t<0.02$ and begin 
to diverge from each other at greater values of $t$.
Nevertheless, the divergence between the average  asymmetries $\ds |f_{t_1,V}(\tau)-f_{t_2,V}(\tau)|$
is more than order of magnitude smaller than the 
difference 
$\ds |f_{t_1,V}(t_1)-f_{t_1,V}(t_2)|$ even in the worst case:
\beq 
|f_{t_1,V}(\tau)-f_{t_2,V}(\tau)| <\!\!\!< |f_{t_1,V}(t_1)-f_{t_1,V}(t_2)|
\eeq 
 $\forall \tau$ and $\forall t_1,t_2$ under study.
The latter quantity is related to the variation 
of the asymmetry owing to its temperature dependence 
through the Polyakov loop, the former --- 
to its explicit temperature dependence. 
The data on $f_{t=0.0051,V}(\tau)$ are presented in the last column of Table~\ref{tab:asymmetry_SU2}.

Thus we conclude that the description of 
the temperature dependence of the asymmetry 
at the given volume in terms of its dependence 
on the Polyakov loop is reliable.

\begin{figure*}[tbh]
\vspace*{-17mm}
\hspace*{-5mm}\includegraphics[width=9cm]{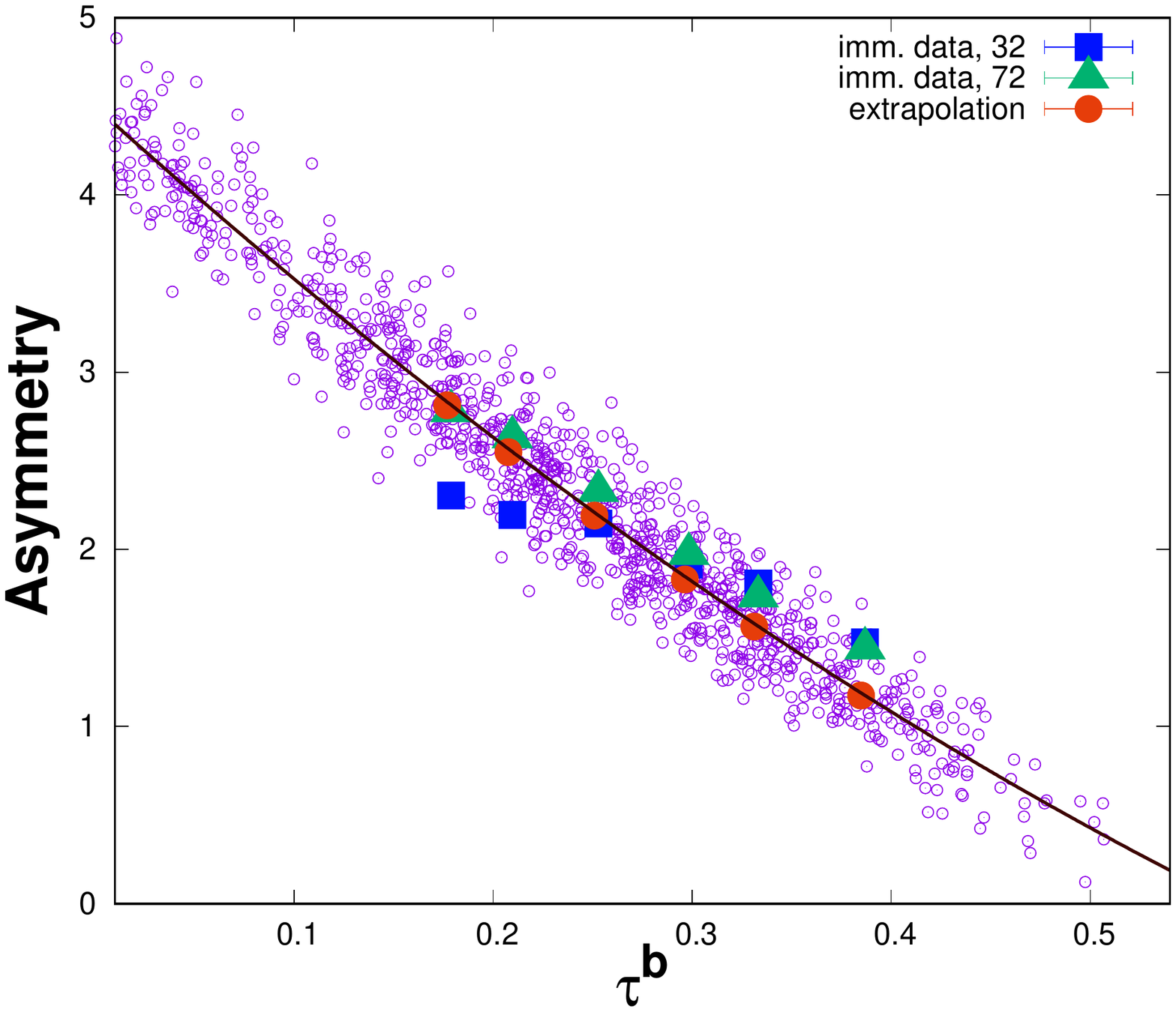}\hspace*{-8mm}
\includegraphics[width=9cm]{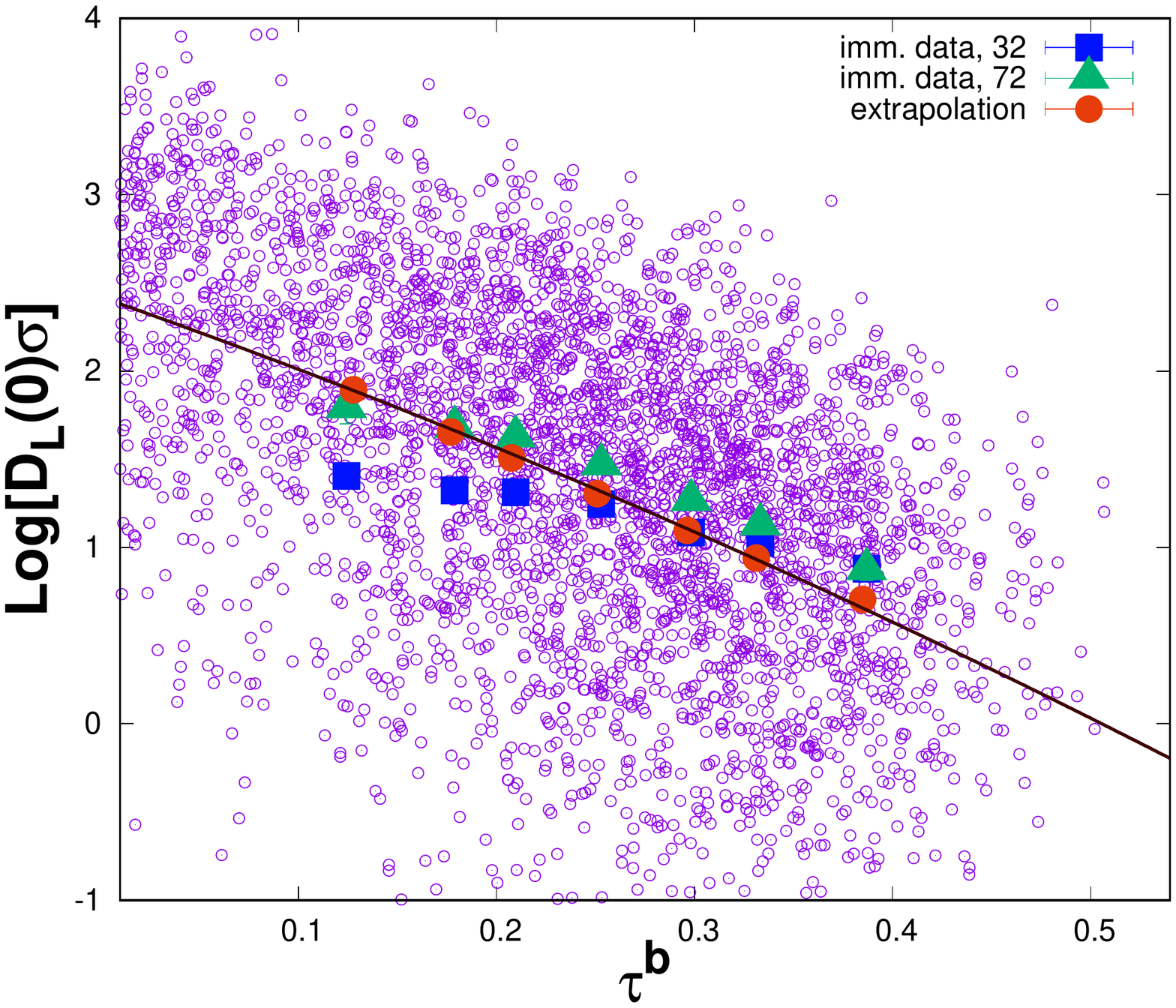}
\vspace*{-35mm}
\caption{ Asymmetry ${\cal A}(\tau)$ (left panel) 
and the logarithm of the zero-momentum longitudinal
propagator ${\cal D}(\tau)$ (right panel) 
extrapolated from the temperature $t=-0.0013$ 
at $L=2.6$~fm are compared with the respective 
averages of the type (\ref{eq:immed_aver}):
squares $L=2.6$~fm and triangles $L=6.0$~fm.}
\label{fig:extrapol}
\end{figure*}

\section{An example of extrapolation from subcritical to supercritical temperatures}

From the very beginning it should be emphasized 
that we rely on the knowledge of the critical 
behavior of the Polyakov loop,
and work in a finite volume where temperature 
can fluctuate through the range under study. 
Therefore, such extrapolation is not forbidden by basic principles of theory.

We perform transformation of the data obtained 
on the lattice 
at $ t=-0.0013$ and $L=2.6$~fm as follows:
\beq \label{eq:data_transform}
(t,{\cal P},{\cal A},{\cal D})\to (\mathbf{p}_V^{-1}({\cal P}),{\cal P},{\cal A},{\cal D})\;.
\eeq 
and arrive at the scatter plot shown in 
Fig.\ref{fig:extrapol}. The average values of 
the type (\ref{eq:improved_asym}) shown by 
solid curves are obtained by regression analysis 
with the use of the bootstrap technique 
(bootstrap errors within orange circles are not seen). 
It is demonstrated that the  values extrapolated 
from the data for a small 
lattice size tend to approach the infinite-volume limit
at $\tau< 0.02$, deviations from this trend at 
larger values of $\tau$ can be explained by 
the need to take into account non-leading 
contributions in formula (\ref{eq:crit_index_beta}).

\section{Conclusions}

We have studied the temperature 
dependence of the chromoelectric-chromomagnetic 
asymmetry and the zero-momentum longitudinal  
gluon propagator at two different 
lattice sizes (2.6~fm and 6.0~fm)
taking into account their correlation with the Polyakov loop.

Our estimates are approximate, however,
they give clear evidence for the following:

\begin{itemize}
\item Finite-volume effects can be significantly
decreased by performing data transformation 
(or improvement of the data) 
(\ref{eq:data_transform}).
\item The temperature dependence of 
the asymmetry ${\cal A}$ and the logarithm 
${\cal D}$ of the dimensionless zero-momentum gluon propagator at the reduced temperature $\tau<0$.
\item The data
on the quantities under study 
at the reduced temperature $\tau<0$ 
involve information on their behavior
at $\tau>0$.  The extrapolation 
of the data related to a finite volume is possible through the range of temperature fluctuations. 
\item Correlations of fluctuating quantities and their consequences deserve more detailed studies 
with regard for finite-size scaling and corrections
to the asymptotic critical behavior.
\end{itemize}

\vspace*{2mm}

\acknowledgments{Computer simulations were performed on the IHEP (Protvino)
Central Linux Cluster and ITEP (Moscow) Linux Cluster.
This work was supported by the Russian Foundation for Basic Research, 
grant~no.20-02-00737~A.}

\bibliographystyle{apsrev}
\bibliography{citations_asym_2023}

\end{document}